# Shape transition of giant liposomes induced by an anisotropic spontaneous curvature

O. Sandre[1,4], C. Ménager[2,*], J. Prost[2], V. Cabuil[2], J-C. Bacri[1], A. Cebers[3]

[1] Laboratoire Milieux Désordonnés et Hétérogènes, UMR 7603 Centre National de la Recherche Scientifique / Université Pierre et Marie Curie, Université Denis Diderot,

Tour 13, Case 78, 4 place Jussieu, 75252 Paris Cedex 05, France.

[2] Laboratoire Liquides Ioniques et Interfaces Chargées, UMR 7612 Centre National de la Recherche Scientifique / Université Pierre et Marie Curie,

Bâtiment F, Case 63, 4 place Jussieu, 75252 Paris Cedex 05, France.

[3] Institute of Physics, University of Latvia, LV-2169 Riga, Salaspils-1, Latvia.

[4] New address: Laboratoire Physico-Chimie Curie, UMR 168

Centre National de la Recherche Scientifique / Institut Curie,

11 rue Pierre et Marie Curie, 75248 Paris Cedex 05, France.

We explore how a magnetic field breaks the symmetry of an initially spherical giant liposome filled with a magnetic colloid. The condition of rotational symmetry along the field axis leads either to a prolate or to an oblate ellipsoid. We demonstrate that an electrostatic interaction between the nanoparticles and the membrane triggers the shape transition.

---

[*] to whom correspondence should be addressed. Telephone number: (33)144272757, Fax number: (33)144273675 and e-mail address: menager@ccr.jussieu.fr.

Phospholipidic bilayers are the basic structural element of most biological membranes. In water these bilayers can form closed vesicles also called liposomes. Due to this analogy with cells, liposomes are commonly used in biophysical studies. In spite of the obvious simplification it implies, artificial liposomes are used to mimic some biological phenomena in well defined, low component environments [1]. For example, analogies have been found between shapes of red blood cells (discocytes, stomatocytes, echinocytes, spherical ghosts) and shapes of synthetic liposomes. Most shapes have a spherical topology and an axis of rotational symmetry: spheres, prolate ellipsoids, dumbells, pear shapes, oblate ellipsoids, discocytes, stomacytes. Other shapes have a symmetry axis of finite order (starfishes), no symmetry axis (3 axed ellipsoids), or another topology (torus). The transformations between those shapes can be triggered by an external parameter: temperature [2, 3], gravity [4], a pH gradient [5, 6] or a photochemical reaction [7]. The area-differential-elasticity model (ADE) enables the attribution of all morphologies to domains in a single phase diagram [8]. The two control parameters are: (i) the spontaneous curvature $c_0$ which reflects the asymmetry between the two sides of the bilayer ; (ii) the 'reduced volume' $v$ which quantifies the ratio of volume to area (high $v$ for an inflated vesicle, low $v$ for a flaccid one). It was shown in [9] that a vesicle can move from one equilibrium shape to another by subtle changes of $c_0$, i.e. by perfusion of different sugars (in osmotic equilibrium). In reference [10] the authors theoretically consider asymmetric bilayers in contact with colloids, but they do not include any angular anisotropy.

In this article, we describe an anisotropic spontaneous curvature due to magnetic nanoparticles placed in the interior of giant liposomes. An aqueous solution of those particles is an ionic ferrofluid [11]. The liposomes filled with such a ferrofluid are called 'magnetoliposomes'. Here, we observe magnetoliposome shapes under a magnetic field of low intensity (400 Oe). In some cases liposomes elongate in the direction of the magnetic field (Fig. 1(a)), while in others they are compressed at their poles (Fig. 1(b)). In the following, we describe the relevant experimental parameters determining the shape: the concentration of magnetic nanoparticles inside the liposome ($C_p$), the salt concentration of the medium ($C_s$), the radius of the liposome ($R_0$) when spherical (with no field) and its axes $a$ and $b$ under deformation. Qualitatively, oblate deformations occur for small, weakly magnetized liposomes, at low ionic strengths. On the other hand, prolate deformations are observed for large, highly magnetized liposomes in salty media. In the last section we propose a model to explain this transition between prolate and oblate shapes.

**Preparation of the magnetoliposomes**

The phospholipid constituting the membrane is 1-2-dioleoyl-sn-glycero-3-phosphocholine (DOPC) purchased as powder from Avanti Polar Lipids. The process used to encapsulate the ferrofluid inside the liposomes with maximal efficiency differs slightly from usual ones: the first step is to pre-hydrate the phospholipid film with the colloidal solution of ferrofluid, and then swell it with pure water. A small amount of perfectly dry powder (around 1mg) of DOPC is mixed with 10µL of the aqueous dispersion (pH 7) of magnetic nanoparticles ($\gamma Fe_2O_3$) and spread and sheared with a glove finger on a glass support (Petri dish) to obtain an oily orange film. This film is presumably a lamellar phase swelled with charged particles. Immediately following the shearing, 1 mL of distilled water is poured onto the film of fat to start the spontaneous swelling of liposomes. Samples are placed in a water bath at 45°C and observed after half an hour with an optical microscope (Leica 40x, NA 0.65). Pictures from a CCD camera are digitized with a frame-grabber (LG-3, Scion Corp., Frederick MD). Most of the liposomes prepared this way are quasi-spherical with diameters ranging from 10 to 100 µm. Their interiors are orange-colored and their membranes exhibit thermal fluctuations.

**Measurement methods**

There is heterogeneity in the colloid entrapment yields of different liposomes. First, the film pre-hydrated with the colloid is mixed manually and hence the particles are not dispersed homogeneously. In addition, liposomes in the same sample can also differ in their 'histories': their membranes close at different times during the swelling process, which competes with the simple diffusion of the ferrofluid by water (without entrapment in a closed membrane). Thus, the concentration $C_p$ of magnetic nanoparticles encapsulated in a given liposome must be measured independently. Therefore we make magnetophoresis experiments. Briefly, the method consists in applying a controlled magnetic field gradient. The intensity of the field gradient is 100 Oe cm$^{-1}$, and the geometry of the field lines is described in detail in [12, 13]. Because magnetoliposomes have a higher magnetic susceptibility $\chi$ than the surrounding liquid, they move towards increasing field intensity. Their velocity is constant and corresponds to the balance of the magnetic force and the drag force exerted by the outer fluid. For ellipsoidal shapes the two forces have analytical expressions, which allows a precise measurement of $\chi$. To get the particle concentration $C_p$, we use the ferrofluid property of superparamagnetism: the susceptibility is the product of $C_p$ and the susceptibility per particle $\chi_p = \dfrac{V_p^2 m_s^2}{3kT}$. The magnetization at saturation of $\gamma Fe_2O_3$, $m_s$=360 Oe, and the average volume of a nanoparticle $V_p$ are known from former studies [14].

Maghemite particles have diameters of the order of 10 nm. Their surface is coated by citrate ligands so that they bear negative charges at pH 7: this is essential for the colloidal stability of the ferrofluid. With no precaution, the concentration of tri-sodium citrate in ferrofluids is rather high ($C_s^0 \approx 90$ mM) because of the non adsorbed citrate species in solution in equilibrium with the citrate ligands. This can be decreased by dialysis through a cellulose membrane (Spectra/Por MWCO 12-14000, ROTH, France). With this method, the salt concentration $C_s^0$ is reduced to 3 to 80 mM as obtained from the measured conductivity of the ferrofluid. In the same way as the nanoparticles, the electrolyte is also diluted in the vesicles compared to the initial ferrofluid. We make the assumption that the sodium citrate salt and the nanoparticles have identical entrapment yields:

$$\frac{C_s}{C_s^0} = \frac{C_p}{C_p^0}$$

. By contrast, if the sodium and citrate ions were not entrapped as efficiently as the particles (e.g. due to their faster diffusion before the bilayers close), then the electrolyte would be homogeneously diluted when adding pure water. For instance, starting with a ferrofluid of $C_s^0$=80 mM would lead to magnetoliposomes in a liquid bath with a salt concentration $C_s$=0.8 mM. The suspension of magnetoliposomes would then have a moderate electrical conductivity ($\approx$1mS cm$^{-1}$ predicted). This is not observed experimentally, as the measured conductivity is less than 40 μS cm$^{-1}$. Therefore we conclude that the electrolyte is indeed co-encapsulated with the magnetic colloid in the liposome interiors.

Dimensions of magnetoliposomes are measured by optical bright field microscopy using image analysis software (Image, NIH). It is well known that the projected area of giant liposomes is smaller than the true surface of the membrane [15], the difference being absorbed by thermal undulations. The effect of a moderate tension is to flatten the fluctuations. Quasi-spherical vesicles can change their shape at constant volume by using the excess area hidden in the fluctuating membrane. In our case, tension originates from magnetic polarization forces acting on the magnetoliposomes. The shape of a liposome submitted to a magnetic field can be described by an ellipsoid with axial symmetry around the field direction. A slight deviation from the rotational symmetry is observed by confocal microscopy (Fig. 2). Giant liposomes filled with a ferrofluid have indeed a higher mass density than the suspending bath: $\Delta\rho \approx 10^{-4} - 5\times 10^{-3}$ g cm$^{-3}$. Thus they are somehow flattened by gravity $\vec{g}_0$ in the $z$ direction. In the next section we develop a simple theory that ignores gravity by considering only axisymmetric ellipsoids. In a standard approximation, shapes are predicted in the zero gravity limit, and gravitational corrections are considered afterward [8, 16]. The dimensionless number $g = \dfrac{g_0 \Delta\rho R_0^4}{K_b}$ has been introduced in [4] to predict deformations of vesicles due to gravity ($K_b$ is the membrane bending modulus). Magnetoliposomes have $g$ values between $10^{-3}$ and 1. This is why the largest ones are not perfectly axisymmetric. Nevertheless, we can clearly distinguish between prolate and oblate shapes, either elongated or flattened in the $y$ direction (parallel to the applied magnetic field). We also point out here the discrepancy of our case with spontaneous transitions between a prolate and an oblate shape reported in [16]. In the latter case, the axis of rotational symmetry also flickers from parallel to perpendicular to the bottom surface of the observation chamber. In the shape transition triggered by a magnetic field, both prolate and oblate ellipsoids keep their axis parallel to $\vec{H}$. Let $a$ be the semi-axis parallel to the magnetic field $\vec{H}$ and $b$ the value of the two other semi-axes perpendicular to $\vec{H}$. The two types of deformations and their amplitudes are characterized by a single parameter of ellipticity: $e^2 \equiv 1 - \frac{b^2}{a^2}$. This parameter is positive for an elongated liposome (*prolate* ellipsoid) and negative for a compressed one (*oblate* ellipsoid).

In summary, the following experiments have been performed. First, we synthesize different ferrofluids with fixed particle concentration $C_p^0$, and varying ionic concentrations $C_s^0$. Then we encapsulate these ferrofluids in giant vesicles, thereafter characterized by optical microscopy. We measure their radius $R_0$ in the quasi-spherical state. Then we use a home built setup of magnetophoresis to simultaneously measure the ellipsoidal deformation under a magnetic field (value of $e^2$ and its sign), and the velocity under the field gradient (perpendicular to the field direction). As explained above, with these measurements we can calculate the concentrations $C_p$ and $C_s$ in individual liposomes.

**Theory**

The overall shape of a liposome subjected to a field $H_0$ must minimize the total free energy, which is the sum of the magnetic energy $E_m$, the bending energy $E_b$, and the surface energy of the vesicle $\tau(S - 4\pi R_0^2)$, where $\tau$ is the membrane tension, and $S$ the surface area of the vesicle. The idea is to calculate the free energy of an initially spherical liposome deformed at constant enclosed volume into an axisymmetric ellipsoid. In the expansion of the free energy as a function of the ellipticity, the terms of order higher than $e^2$ are neglected. The surface energy thus disappears since it varies as $e^4$.

The magnetic term is proportional to the liposome volume. It is related to the shape through the demagnetizing factor $D(e)$:

$$E_m = -\frac{2\pi}{3} \frac{\chi H_0^2}{1 + 4\pi\chi D(e)} R_0^3 \qquad (1)$$

where $\chi$ is the magnetic susceptibility of the encapsulated liquid. Using a standard expansion of $D$ as a function of $e$ [17], one obtains the approximate expression:

$$E_m \approx \text{constant} - \frac{(4\pi\chi H_0)^2 R_0^3}{45} e^2 \qquad (2)$$

This magnetic energy is decreased by positive values of $e^2$, and thus it favors a prolate deformation.

In the following we show that the prolate/oblate transition under magnetic field is driven by a competition between the magnetic energy $E_m$ and the bending energy $E_b$. The strong influence of the ionic strength on the deformation suggests that electrostatic interactions are involved between the charged nanoparticles and the slightly charged membrane. We make here the assumption that some charges exist on the DOPC phospholipidic bilayers. This assumption is supported by previous studies which report: i) the existence of a weak negative $\zeta$-potential of a few mV for vesicles made of natural egg lecithin [18] or synthetic phosphatidylcholine lipids [19]; and ii) the evidence of electrostatic repulsion between pure phosphatidylcholine bilayers, measured with a Surface Force Apparatus [20].

The behavior of charged membranes has been studied extensively in theoretical papers [21, 22, 23]. The electrostatic contributions to the bending modulus $K_b$ and to the spontaneous curvature $c_0$ have been considered in [24] in the Debye-Hückel approximation of the electrostatic potential and in the case of electrically uncoupled leaflets. It appears that electrostatic repulsion produces an enhancement of the bending modulus compared to the uncharged case: surface charges make the bilayer stiffer. The spontaneous curvature $c_0$ characterizes the bilayer asymmetry [25]. It has been derived in reference [24] for a bilayer bearing different surface charge densities on its two sides. In that case the salt concentrations were identical inside and outside the vesicle, and hence the Debye lengths as well. Here we develop another model that fits better to the case of magnetoliposomes: asymmetry comes from two different screening lengths $\kappa_i^{-1}$ and $\kappa_o^{-1}$, inside and outside respectively.

In a suspension of magnetoliposomes, the outer phase is a usual salt solution, while the inner phase is a complex colloidal system combining the salt and the charged magnetic nanoparticles. In order to obtain a simple solution of this problem, we describe this inner phase as a mixture of electrolytes: – the particles (concentration $C_p$), identified as magnetic polyions with an effective charge $Z$ – the other ions (charge $Z_i$ for type $i$) having the same concentration $C_s^i$ as in the suspending bath. In the outer phase, electrostatic interactions are screened by the ions (in practice sodium and citrate, but no particles) above a standard isotropic Debye length

$$\kappa_o^{-1} = \left( \frac{\varepsilon_w kT}{4\pi \sum_i C_i Z_i^2 e^2} \right)^{1/2}$$

, where $\varepsilon_w$ is the dielectric constant of water ($\varepsilon_w$=78.5). Inside the liposomes, the decay length of electrostatic interactions is denoted $\kappa_i^{-1}$ and takes into account all the charged species, salt ions and nanoparticles. The modifications on the bending modulus and on the spontaneous curvature are estimated using the same approach as in [24]. At first we calculate the electrostatic energy $E_{el}$ of a simple body – an infinite cylinder with radius $R$ – then we expand it as a function of $R$. Finally we identify the terms with the standard expression of the bending energy, supposing that these expressions remain the same for a body of general geometry:

$$\frac{E_{el}}{2\pi R} \approx \frac{1}{2} \frac{4\pi \sigma^2}{\varepsilon_w} \left( \frac{1}{\kappa_o} + \frac{1}{\kappa_i} + \frac{1}{2\kappa_i^2 R} - \frac{1}{2\kappa_o^2 R} + \frac{3}{8\kappa_i^3 R^2} + \frac{3}{8\kappa_o^3 R^2} \right) \equiv \frac{1}{2} K_b^{el} \left( \frac{1}{R} - c_0 \right)^2 \quad (3)$$

The elastic constants are derived in the assumption of a slight difference between $\kappa_i$ and $\kappa_o$. The bending modulus is then $K_b^{el} \cong \frac{3\pi \sigma^2}{\varepsilon_w \kappa_o^3}$ and the spontaneous curvature becomes:

$$c_0 = \frac{\kappa_o^3}{3} \left( \frac{1}{\kappa_o^2} - \frac{1}{\kappa_i^2} \right) \quad (4)$$

Here $\sigma$ denotes the surface charge density. It is inferred that $\sigma$ is the same on both monolayers, because the pH is buffered inside and outside by the citrate salt. The asymmetry between the two sides of the membrane arises from different electrolyte compositions inside and outside the vesicle. This small difference has a negligible effect on the modulus, but it is dominant for the spontaneous curvature.

In the model of a liposome filled with a ferrofluid, we make an additional approximation. We suppose that the difference between $\kappa_i$ and $\kappa_o$ is due only to the presence of nanoparticles inside, and their absence outside. In other words, we neglect the difference of salt concentrations inside and outside, responsible for a classical isotropic $c_0$. The model focuses on the field dependence of the inner Debye length $\kappa_i^{-1}$. The concentration profile of nanoparticles near the inner leaflet of the membrane is indeed sensible to $\vec{H}$. Thus, a magnetic term has to be added in the linearized Poisson-Boltzmann equation. Due to that the calculation of the electrostatic energy of the membrane in the presence of the magnetic field turns out to be rather complicated problem since simultaneous solution of the electrostatic and magnetic field equations is involved. As result at the curvature expansion of the electrostatic energy of the membrane its anisotropy arising due to the magnetic interactions appears. This electrostatic energy curvature expansion is carried out in the forthcoming publication [27]. Its results confirm the conclusions of much simpler model described below. The simple model is based on the assumption of the locally flat membrane and the reduction of the solution of the coupled electrostatic and magnetostatic field equations to 1D approximation. Obtained in such way screening constants are substituted in the expressions of the bending elasticity constant and spontaneous curvature leading to the anisotropic curvature elasticity energy of the membrane. Further consideration is carried out in the frame of the simple model. We begin by solving the magnetostatic problem in a flat membrane geometry, with a magnetic field at an angle $\alpha$ with the normal to the plane of the membrane. Let $x$ be the coordinate in the direction normal to the membrane. Because the vesicle radius is three orders of magnitude larger than the particles sizes, we can neglect the membrane curvature and the finite size of the nanoparticles. Thus we denote the nanoparticle concentration $C_p(x)$ at a distance $x$ from the membrane, the $x$ component of the field $H_x(x)$, and the electric potential $\psi(x)$. Far away from the membrane those variables tend respectively towards $C_p$, $H_0 cos(\alpha)$ and zero. The solution of Maxwell equation $div(\vec{H} + 4\pi\vec{M}) = 0$ is:

$$H_x(x) = H_0 cos(\alpha) \frac{1 + 4\pi C_p \chi_p}{1 + 4\pi C_p(x) \chi_p} \qquad (5)$$

The net flux of nanoparticles in the $x$ direction is then calculated using their mobility $u$ under an osmotic, an electrophoretic and a magnetophoretic force:

$$J_x = -u C_p(x) Z \frac{\partial \psi}{\partial x} - u k T \frac{\partial C_p(x)}{\partial x} + u C_p(x) \chi_p H_x \frac{\partial H_x}{\partial x} \qquad (6)$$

At equilibrium, $J_x=0$ and:

$$kT \left[ 1 + 4\pi \frac{C_p(x) \chi_p^2 (1 + 4\pi C_p \chi_p)^2 H_0^2 cos^2(\alpha)}{kT (1 + 4\pi C_p(x) \chi_p)^3} \right] \frac{\partial C_p(x)}{\partial x} = -C_p(x) Z \frac{\partial \psi}{\partial x} \qquad (7)$$

In the Poisson-Boltzmann theory and the Debye Hückel approximation, a solution is found for a small variation of nanoparticle concentration around the bulk value ($\delta C_p(x) = C_p(x) - C_p$, $\delta C_p(x) << C_p$). Then the linear dependence of $\delta C_p(x)$ with the electrostatic potential $\psi(x)$ becomes anisotropic,

$$\delta C_p(x) = C_p \frac{-Z\psi(x)}{kT}\left[1 + \lambda \cos^2(\alpha)\right]^{-1} \quad \text{where} \quad \lambda = \frac{4\pi\chi^2 H_0^2}{C_p k_B T(1 + 4\pi\chi)}$$

compares the two pressures acting on the membrane: the magnetic pressure calculated at the liposome poles ($2\pi\chi^2 H_0^2$) and the osmotic pressure exerted by the nanoparticles ($C_p kT$). Numerical values of $\lambda$ range from $10^{-3}$ to $5\times10^{-2}$, as seen in Table 1. Looking back to the electrostatic potential $\psi(x)$, an anisotropic decay factor can be defined, including the screening by all the charged species:

$$\kappa_i(\alpha) = \kappa_o\left[1 + \frac{d}{1 + \lambda \cos^2(\alpha)}\right]^{\frac{1}{2}} \quad (8)$$

It depends through $d = \dfrac{Z^2 C_p}{\sum_i Z_i^2 C_s^i}$ on the ionic strength of the dispersion and through $\lambda$ on its magnetic characteristics. The anisotropic part of $\kappa_i$ is negligible as long as the screening is dominated by the salt ions, i.e. for high ionic strengths (small $d$). In the opposite case, the angular dependence of the distribution of particles induces an angular dependence of $\kappa_i(\alpha)$, and hence of $c_0(\alpha)$. The largest screening length turns out to be on the poles. The reason for that is rather simple. Since the magnetic pressure acting near the poles pushes the magnetic polyions to the membrane it augments near it the concentration of the charges having the same sign as on the membrane and thus diminishes the ecranization of charges on it. An increase of the screening length near the magnetic poles ($\alpha=0$ and $\alpha=\pi$) decreases locally the spontaneous curvature: qualitatively, the oblate shape is favored because it has a lower curvature at the poles than at the equator. Equations (4) and (8) enable one to derive the spontaneous curvature at any angle $\alpha$:

$$c_0(\alpha) \approx \frac{\kappa_o d}{3(1+d)}\left[1 - \frac{\lambda}{1+d}\cos^2(\alpha)\right] \quad (9)$$

We can now calculate precisely the bending energy of the membrane, $E_b = \dfrac{1}{2}\left(K_b + K_b^{el}\right)\int\left[\dfrac{1}{R_1} + \dfrac{1}{R_2} - c_0(\alpha)\right]^2 dS$. It involves the two local principal radii of curvature $R_1$ and $R_2$ of the bilayer [25]. Due to the dependence of the bending elasticity modulus on the screening constant and its anisotropy described by the relation (8) anisotropic under the action of the magnetic field, the bending elasticity modulus of the membrane is also anisotropic. Nevertheless its account is not necessary at the calculation of the curvature elasticity energy of the vesicle up to the first order terms in $e^2$ since the direct calculation shows that its contribution in that approximation disappears exactly. That is why for the bending elasticity modulus the expression $K_b^{el} = \dfrac{3\pi\sigma^2}{\varepsilon_w \chi_0^3}$ is assumed, what does not reflect the dependence of the modulus on the magnetic field strength and its direction. As before, we look at an ellipsoidal deformation of a quasi-spherical vesicle. Integration leads to:

$$E_b \approx \frac{32\pi(K_b + K_b^{el})}{405} \lambda d^2 (1+d)^{-3} (\kappa_o R_0)^2 e^2 \qquad (10)$$

Exact treatment of the electrostatic energy of the vesicle under the action of the magnetic field carried out in [27] shows that besides the anisotropies of the elastic modulus discussed above, there is specific 3D efect leading to the anisotropy of the bending modulus depending on the tangential component of the magnetic field strength. Howewer its contribution to the electrostatic energy is negligible since it does not contain – in comparison with the expression (10) – a multiplier $(\chi_0 R_0)^2$ which, as it follows from data in Table 1, is very large.

$E_b$ enters in the expansion of the total free energy with the same order in $e^2$ as the magnetic energy $E_m$, but with the opposite sign. Indeed it is possible to explain a prolate deformation or an oblate one depending whether $E_m$ or $E_b$ is predominant.

**Results and discussion**

The experimental measurements of $R_0$ and $\chi$ on given liposomes are computed in Eq. (1) to get the slopes in the expansion of $E_m$ as a function of $e^2$. Then we convert the values of $\chi$ into particle concentrations $C_p$, inferring a mean value for the nanoparticles diameters $D=12\times10^{-7}$ cm, or equivalently a magnetic susceptibility per particle $\chi_p=8.6\times10^{-19}$ cm$^3$. The salt concentration is deduced by $C_s=C_s^0 C_p/C_p^0$. From stoichiometry it follows that $Z_1=3$ and $C_1=C_s$ for the citrate anions, $Z_2=1$ and $C_2=3C_s$ for the sodium cations. Then we calculate the electrostatic decay factor $\kappa_o$ caused by the salt ions only. The two parameters $\lambda$ and $d$ appearing in the anisotropic term of $\kappa_i(\alpha)$ are computed from the values of $\chi$, $C_p$, $C_s$ and $Z=25$, a reasonable estimate of the net charge per particle (including the adsorbed counterions). The surface charge density of the bilayer is inferred to be $\sigma=800$ ues cm$^{-2}$. It represents about 1% of negatively charged lipids. Finally we compute $E_b/e^2$ from Eq. (10). The ratio of the two competing terms $E_b/E_m$ is plotted vs. $C_s$ on Fig.3. For the higher ionic strengths where the screening of charges is the strongest, the magnetic term dominates the electrostatic contribution to the bending energy: the deformation is indeed prolate. The field dependent elongation has been analyzed in a former work in order to determine the bending modulus of the DOPC bilayer, $K_b=(8.6\pm0.8)\times10^{-13}$ dyn cm [26]. But a dilution of the ions in the bulk by two orders of magnitude is sufficient to reverse the ratio of $E_b$ and $E_m$, thus explaining the oblate shape. The critical value for this ratio is 1, and it clearly separates the graph into domains of prolates and oblates. The scatter of Fig. 3 is due to two parameters, other than $C_s$, that vary in a polydisperse suspension of magnetoliposomes: the vesicle radius $R_0$ and the encapsulated particle concentration $C_p$ (or equivalently the $\lambda$-parameter). For example the magnetic energy $E_m$ varies like $R_0^3$ and the anisotropic bending energy $E_b$ like $R_0^2$. The scatter of the data has been reduced by plotting on Fig. 3 the ratio $E_b/E_m \propto R_0^{-1}$. The error bars estimate the uncertainty in our measurements (shape analysis, magnetophoresis and conductimetry). It is impossible to fit all the data with a single simulation of Eqs. (1) and (10) because of the polydispersities of $R_0$ and $\lambda$. This gives the residual scatter in the data. However the sensitivity on $C_s$ is fully established: $C_s$ is varied over two orders of magnitude, and this inverts the weights of $E_b$ and $E_m$ in the total energy. The set of values for the guessed parameters – $Z$, $\chi_p$ (or equivalently $D$) and $\sigma$ – is reasonable. The point that needs to be further discussed in our model is the approximation of a flat interface to derive the magnetic field $H_x(x)$ as given by Eq. (5). Complete treatment of the problem based on the curvature expansion of the electrostatic energy of the membrane of the arbitrary shape is carried out in [27]. Results obtained completely confirms qualitative picture considered above. It should be remarked that results obtained for the anisotropy of the spontaneous curvature in the frame of the complete treatment shows the coincidence with qualitative simple model even up to the numerical coefficient. Therefore current model is sufficient to point out that the electrostatic double layer of membrane is deformed by the magnetic field *via* its effect on the charged magnetic nanoparticles.

We have reported an original shape transition of giant magnetic liposomes triggered by the ionic strength. The transition between prolate and oblate ellipsoids has already been observed under an applied ac *electric field* [28]: the vesicles are prolate in the 'conductive regime' (low frequency) and oblate in the 'dielectric regime' (high frequency). It has also been predicted for an applied *magnetic field*: Helfrich showed that the sign of magnetic anisotropy of the lipid molecules themselves leads either to prolate or oblate deformations [25, 29]. However the magnetically induced ellipticity has been far too small to be detected experimentally, even in a strong field $H_0=10^4$ Oe (1 Tesla). Here we have taken advantage of the magnetic properties of the encapsulated solution to improve the coupling between the magnetic field and the bending energy of the vesicles. The main result of our theoretical description is an anisotropic distribution of particles at the neighborhood of the membrane when a field is applied. Both the Debye screening length at the inner monolayer and the spontaneous curvature of the bilayer maintain this anisotropy. This model corresponds to the regime of electrostatic interaction between a weakly charged phospholipid bilayer and ionic encapsulated species. This might be an important point for the description of biological membranes which are surrounded by charged mesoscopic objects and embedded in salty media.

**Acknowledgments**

We thank J. Mertz and E. Karatekin for reading the manuscript and M. Blanchard-Desce for the gift of Di-6-ASPBS.

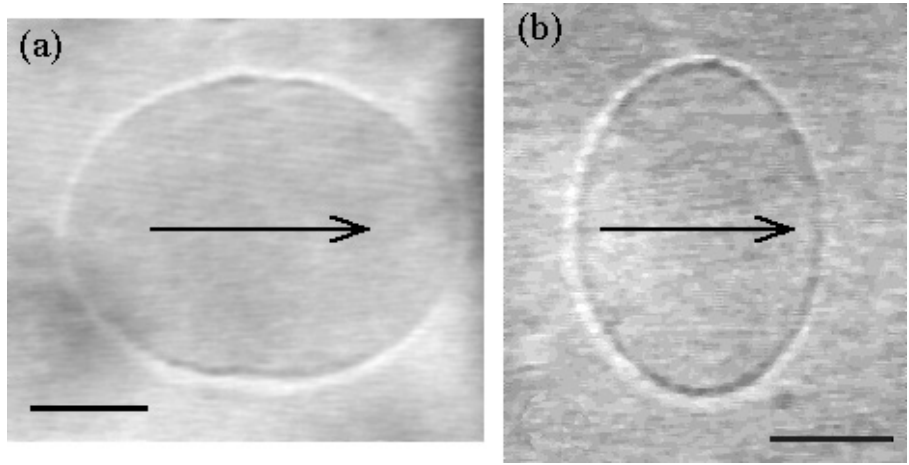

**Fig. 1**: The two competing shapes of magnetoliposomes subjected to a magnetic field (amplitude $H_0$=400 Oe, field direction along the arrow); (a) prolate ellipsoid for a high ionic strength ferrofluid ($C_s^0$=85 mM); (b) oblate when the ionic strength has been lowered by dialysis ($C_s^0$=7.75 mM). Length of the bar is 10 µm.

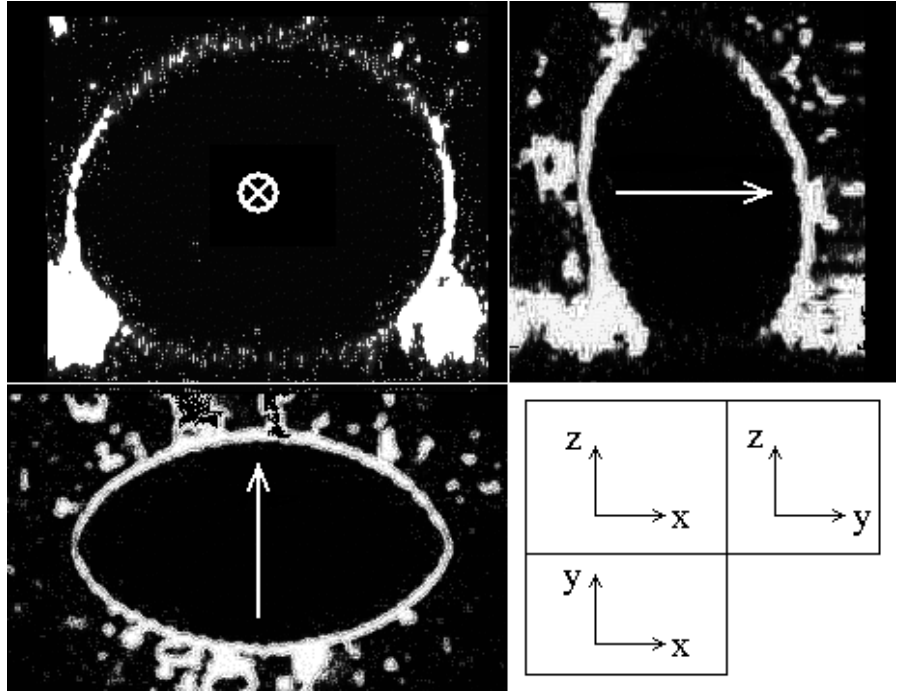

**Fig. 2**: Cross-sections of an oblate magnetoliposome in the three principal planes *xy*, *xz* and *yz*. A stack of 150 pictures in the sample plane (*xy*) is taken every 0.5μm along the *z* axis with a confocal microscope (TCS4D, Leica). The lipid bilayer is labelled with the fluorescent dye Di-6-ASPBS (N-(4-sulfobutyl)-4-(4-(dihexylamino)styryl)pyridinium). Excitation is provided by the 488nm line of an Argon laser, and fluorescence is collected through a fluorescein filter set. The ellipsoid semi-axes are *a*=19μm and *b*=32.5μm. The ferrofluid has a low ionic strength ($C_s^0$=5 mM). The white arrow indicates the magnetic field (intensity $H_0$=200 Oe). Gravity is along the *z* axis. The hot spots are lipid-particle aggregates that either settle at the bottom of the cell or make chains attracted by the strongly magnetized giant liposome.

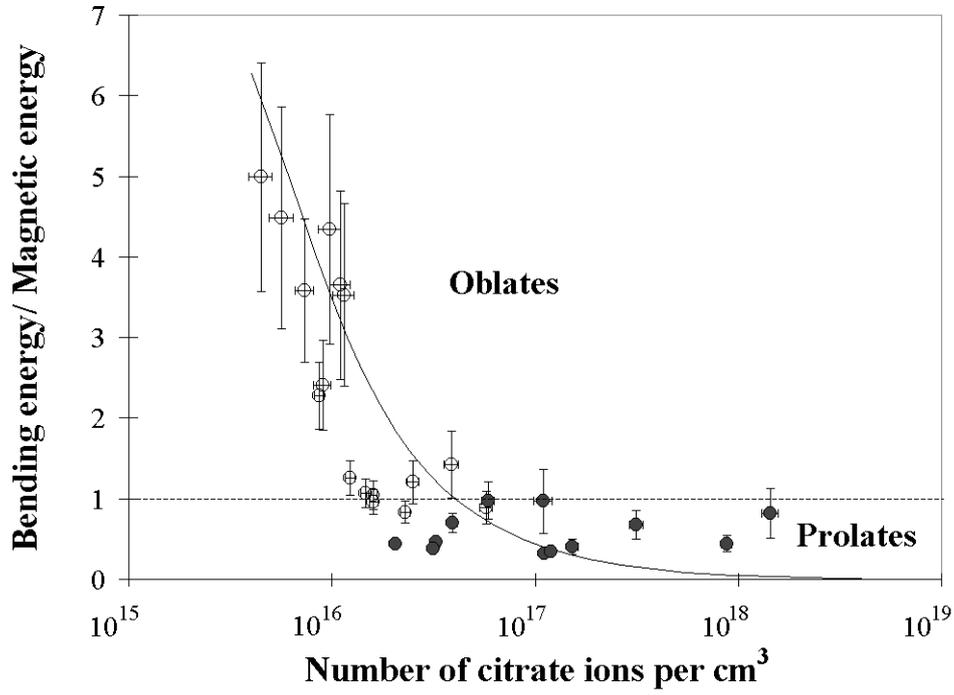

**Fig. 3**: Ratio of the two competing terms $E_b/E_m$ in the expansion of the free energy as a function of ellipticity. They are calculated for oblate (open markers) and for prolate (filled markers) liposomes using both experimental data ($C_p$, $C_s$, $R_0$) and assumed parameters (see text). Solid line is plotted from Eqs. (1) and (10) for typical values $R_0$=15 μm and $\lambda$ =5×10$^{-3}$.

| $R_0$ (μm) | $e^2 = 1 - \dfrac{b^2}{a^2}$ | $d$ | $\lambda$ | $\kappa_o^{-1}$ (nm) | $K_b^{el}/K_b$ | $c_0^{-1}$ (nm) |
|---|---|---|---|---|---|---|
| 25.2 | -0.397 | 0.96 | $3.0 \times 10^{-3}$ | 45 | 8.4 | 280 |
| 13.3 | -0.596 | 0.62 | $4.8 \times 10^{-3}$ | 29 | 2.2 | 230 |
| 21.4 | -1.17 | 0.62 | $3.2 \times 10^{-3}$ | 36 | 4.0 | 280 |
| 24.6 | -0.845 | 0.19 | $3.1 \times 10^{-3}$ | 20 | 0.73 | 380 |
| 22.5 | 0.496 | 0.19 | $5.3 \times 10^{-3}$ | 15 | 0.32 | 290 |
| 50.4 | 0.343 | 0.19 | $2.8 \times 10^{-3}$ | 21 | 0.85 | 400 |
| 21.1 | -1.31 | 0.14 | $1.5 \times 10^{-3}$ | 24 | 1.3 | 600 |
| 31.6 | 0.466 | 0.11 | $2.3 \times 10^{-3}$ | 17 | 0.45 | 540 |
| 14.1 | 0.536 | 0.074 | $4.5 \times 10^{-2}$ | 3.2 | $3.1 \times 10^{-3}$ | 150 |
| 13.0 | 0.586 | 0.056 | $6.1 \times 10^{-3}$ | 7.8 | $4.2 \times 10^{-2}$ | 440 |

**Table 1**: Examples of numerical data for oblates (clear cells) and prolates (shadowed cells). $R_0$ and $e^2$ are measured by optical microscopy. The parameters $d$ and $\lambda$ are calculated from experimental measurements of the concentrations $C_s$ and $C_p$. The last three columns contain important features of the model: the screening length $\kappa_o^{-1}$ due to the salt ions only, the ratio of the electrostatic contribution to the bending modulus of membrane, and the local value of the spontaneous curvature radius at the magnetic poles ($\alpha=0$ and $\alpha=\pi$).